\documentclass[twocolumn,showpacs,preprintnumbers,amsmath,amssymb]{revtex4}

\usepackage{graphicx}
\usepackage{dcolumn}
\usepackage{bm}

\newcommand{\bq}{\begin{equation}}
\newcommand{\eq}{\end{equation}}
\newcommand{\bqa}{\begin{eqnarray}}
\newcommand{\eqa}{\end{eqnarray}}
\newcommand{\nn}{\nonumber \\}

\def\be     {\begin{equation}}
\def\ee     {\end{equation}}
\def\bea        {\begin{eqnarray}}
\def\eea        {\end{eqnarray}}
\def\bnn    {\begin{eqnarray*}}
\def\enn    {\end{eqnarray*}}

\begin{document}

\title{Bandwidth-control vs. doping-control Mott transition in the Hubbard model}
\author{Ki-Seok Kim}
\affiliation{School of Physics, Korea Institute for Advanced
Study, Seoul 130-012, Korea}
\date{\today}

\begin{abstract}
We reinvestigate the bandwidth-control and doping-control Mott
transitions (BCMT and DCMT) from a spin liquid Mott insulator to a
Fermi liquid metal based on the slave-rotor representation of the
Hubbard model,\cite{Florens} where the Mott transitions are
described by softening of bosonic collective excitations. We find
that the nature of the insulating phase away from half filling is
different from that of half filling in the respect that a charge
density wave coexists with a topological order (spin liquid) away
from half filling because the condensation of vortices generically
breaks translational symmetry in the presence of "dual magnetic
fields" resulting from hole doping while the topological order
remains stable owing to gapless excitations near the Fermi
surface. Performing a renormalization group analysis, we discuss
the role of dissipative gauge fluctuations due to the Fermi
surface in both the BCMT and the DCMT.
\end{abstract}


\maketitle

\section{Introduction}

Landau-Ginzburg-Wilson (LGW) paradigm has been our unique
theoretical framework for classical phase transitions. Starting
from an electron Hamiltonian, one can derive an effective LGW free
energy functional in terms of order parameters associated with
some symmetry breaking. The LGW framework has been also applied to
quantum phase transitions by taking temporal fluctuations of order
parameters into account, usually called the Hertz-Millis
theory.\cite{HM}

There are models of quantum phase transitions, on the other hand,
which may defy interpretation in the LGW paradigm. Consider the
superfluid-insulator transition of a boson Hubbard-type model. The
boson density ($n$) and its phase ($\phi$) are canonically
conjugate, satisfying the uncertainty relation $\Delta n \Delta
\phi \gtrsim 1$, and the competing nature of the two variables
results in the condensation of one variable or the other,
depending on the ratio of phase stiffness and the compressibility.
The quantum conjugate nature of the variables, satisfying an
uncertainty relation, lies at the heart of the quantum phase
transition in this particular case. It is not obvious how LGW
theory, written solely in terms of an order parameter, captures
the inherent competing nature of the conjugate variables driving
the quantum phase transition.

As another example of a quantum phase transition where an order
parameter description is likely to fail, we mention the metal to
paramagnetic insulator transition (Mott transition) found in the
study of the two dimensional Hubbard model. As recent dynamical
mean-field theory (DMFT) studies show, the transition is
associated with the vanishing of the spectral weight of the
quasiparticle peak, but not with any symmetry breaking field,
hence the order parameter approach in the LGW paradigm is not
clear to be applicable.\cite{DMFT}

Recently, Florens and Georges (FG) reexamined a bandwidth-control
Mott transition (BCMT) from a paramagnetic Mott insulator of a
spin liquid to a correlated metal of a Fermi liquid at half
filling in the Hubbard model.\cite{Florens} In order to describe
the BCMT they introduced an elegant formulation based on the
slave-rotor representation, and investigated properties of
metallic and insulating phases of the model. Many of the
properties obtained at the mean-field level matched well with the
more sophisticated DMFT calculations.\cite{DMFT} In this
formulation the competing nature of canonically quantum conjugate
variables naturally appears. Within this theoretical framework the
Mott transition is understood by softening of bosonic collective
excitations, physically associated with zero-sound modes in a
Fermi liquid. When these bosonic excitations are gapped, a
paramagnetic Mott insulator with charge gap but no spin gap
results, thus called a spin liquid. On the other hand,
condensation of the boson excitations causes a coherent
quasiparticle peak at zero energy, resulting in a Fermi liquid in
the low energy limit.

In the present paper we investigate a doping-control Mott
transition (DCMT) from the spin liquid to the Fermi liquid based
on the slave-rotor representation of the Hubbard model. We find
that the DCMT differs from the BCMT in the respect that the nature
of the Mott insulator and the mechanism of the Mott transition are
different from each other. Hole doping results in a nontrivial
Berry phase term to the boson field, leading to an effective
magnetic field for its vortex field in the dual formulation. It is
shown that this effective magnetic field induces a crystalline
phase of doped holes, coexisting with the spin liquid. On the
other hand, the paramagnetic Mott insulator at half filling is the
same spin liquid, but without any charge orders. We argue that the
doped spin liquid with charge order evolves into the Fermi liquid
via a continuous phase transition.

The present scenario for the DCMT was discussed before, but based
on the boson-only (Hubbard) model, where fermionic excitations are
ignored\cite{DHLee,Balents1} or decoupled to the bosonic
excitations\cite{Tesanovic} in the renormalization group (RG)
sense. In this paper we start from the electron Hubbard model, and
derive an effective bosonic field theory. It should be noted that
this effective field theory is totally different from that in the
boson Hubbard model owing to the presence of damped gauge
fluctuations resulting from gapless fermion excitations. Although
the Berry phase plays the same role in both doped (boson and
fermion) Mott insulators, nature of the Mott transitions would be
different owing to the presence of dissipative gauge fluctuations
in the slave-rotor representation of the electron Hubbard model.
Performing an RG analysis, we show that the dissipative dynamics
of gauge excitations makes both the BCMT and the DCMT in the
electron Hubbard model differ from those in the boson Hubbard
model. There also exists a previous study considering fermion
excitations coupled to bosonic fields.\cite{Baleant2} However,
this study starts from the quantum dimer model, and considers a
valance bond solid instead of the spin liquid with a Fermi
surface. Thus, the fermion excitations in the model are gapped
(for the $s-wave$ pairing case), thus ignored in the low energy
limit. We would like to emphasize that our crystalline phase is
nothing to do with Cooper pairs,\cite{Balents1,Tesanovic,Baleant2}
instead associated with doped holes.\cite{DHLee}

\section{Effective field theory for the Mott transition}

We derive the slave-rotor representation of the Hubbard model in
the path-integral formulation. We note that FG derived it based on
not only the canonical quantization method but also the path
integral formulation. However, we argue that our path integral
derivation more clearly shows the connection between
Hubbard-Stratonovich (HS) fields and rotor variables.

We consider the Hubbard model in two dimensions \bqa && H = -
t\sum_{ij\sigma}c_{i\sigma}^{\dagger}c_{j\sigma} +
U\sum_{i}n_{i}^{2} . \eqa Here $t$ is a hopping integral of
electrons, and $U$ the strength of local interactions. $n_{i} =
\sum_{\sigma}c_{i\sigma}^{\dagger}c_{i\sigma}$ is an electron
density.

A usual methodology treating the Hubbard $U$ term is a HS
transformation. Using the coherent state representation and
performing the HS transformation, we obtain the partition function
\bqa && Z = \int{D[c_{i\sigma},\varphi_{i}]}\exp\Bigl[-\int{d\tau}
\Bigl(\sum_{i\sigma}c_{i\sigma}^{*}(\partial_{\tau} -
\mu)c_{i\sigma} \nn && -
t\sum_{ij\sigma}c_{i\sigma}^{*}c_{j\sigma} +
\sum_{i}(\frac{1}{4U}\varphi_{i}^{2} -
i\varphi_{i}\sum_{\sigma}c_{i\sigma}^{*}c_{i\sigma})\Bigr)\Bigr] ,
\eqa where $\varphi_{i}$ is an order parameter associated with a
charge density wave (CDW), and $\mu$, the chemical potential of
electrons. Physically, the $\varphi_{i}$ field corresponds to an
effective electric potential. In the usual mean-field manner the
CDW order parameter is given by $-i\overline{\varphi}_{i} =
2U\langle\sum_{\sigma}c_{i\sigma}^{*}c_{i\sigma}\rangle$.

Integrating over electronic excitations in Eq. (2) and expanding
the resulting logarithmic term for the effective potential
$\varphi_{i}$, one can obtain an effective LGW free energy
functional in terms of the CDW order parameter $\varphi_{i}$. As
mentioned in the introduction, it is not clear that this LGW
theoretical framework has the competing nature of quantum
conjugate variables because there exists only one CDW order
parameter. One can say that the formulation Eq. (2) is exact, and
thus the LGW framework may be a good starting point. However, an
important point is how to expand the resulting logarithmic term.
The expansion should be approximately performed, and thus one
cannot say validity of the LGW framework for quantum phase
transitions.\cite{Chubukov}

It is clear that the metal-insulator transition is associated with
charge fluctuations. One way controlling charge fluctuations is to
introduce the canonical conjugate variable of the charge density.
Unfortunately, ${\varphi}_{i}$ is not the canonically conjugate
variable of the charge density because it is an effective electric
potential.

We consider the gauge transformation for an electron field \bqa &&
c_{i\sigma} = e^{-i\theta_{i}}f_{i\sigma} . \eqa Here
$e^{-i\theta_{i}}$ is assigned to be an annihilation operator of
an electron charge, and $f_{i\sigma}$ an annihilation operator of
an electron spin. In this paper we call $e^{-i\theta_{i}}$ and
$f_{i\sigma}$ chargon and spinon, respectively.

Inserting Eq. (3) into Eq. (2), we obtain \bqa && Z =
\int{D[f_{i\sigma},\theta_{i},\varphi_{i}]}\exp\Bigl[-\int{d\tau}
\nn && \Bigl(\sum_{i\sigma}f_{i\sigma}^{*}(\partial_{\tau} - \mu -
i\partial_{\tau}\theta_{i})f_{i\sigma} -
t\sum_{ij\sigma}f_{i\sigma}^{*}e^{i(\theta_{i} -
\theta_{j})}f_{j\sigma} \nn && +
\sum_{i}(\frac{1}{4U}\varphi_{i}^{2} -
i\varphi_{i}\sum_{\sigma}f_{i\sigma}^{*}f_{i\sigma})\Bigr)\Bigr] .
\eqa

Performing the HS transformation $(1/4U)\varphi_{i}^{2}
\longrightarrow UL_{i}^{2} + iL_{i}\varphi_{i}$, and shifting
$\varphi_{i}$ into $\varphi_{i} \longrightarrow \varphi_{i} -
\partial_{\tau}\theta_{i}$, we obtain the following expression for
the partition function \bqa && Z =
\int{D[f_{i\sigma},\theta_{i},\varphi_{i},L_{i}]}\exp\Bigl[-\int{d\tau}\nn&&
\Bigl(\sum_{i\sigma}f_{i\sigma}^{*}(\partial_{\tau} -
\mu)f_{i\sigma} - t\sum_{ij\sigma}f_{i\sigma}^{*}e^{i(\theta_{i} -
\theta_{j})}f_{j\sigma} \nn && + \sum_{i}[UL_{i}^{2} +
iL_{i}(\varphi_{i} -
\partial_{\tau}\theta_{i}) -
i\varphi_{i}\sum_{\sigma}f_{i\sigma}^{*}f_{i\sigma}]\Bigr)\Bigr] .
\eqa Integrating over the $\varphi_{i}$ field, one finds $L_{i} =
\sum_{\sigma}f_{i\sigma}^{\dagger}f_{i\sigma}$. In this respect
$L_{i}$ corresponds to the density variable of FG.\cite{Florens}

Eq. (5) has an interesting structure for the quantum phase
transition. First of all, there is the competing nature of
canonically conjugate quantum variables. The $\theta_{i}$ field is
canonically conjugate to the charge density $L_{i} =
\sum_{\sigma}f_{i\sigma}^{\dagger}f_{i\sigma}$, as one can see
from the coupling term $-iL_{i}\partial_{\tau}\theta_{i}$ of the
Lagrangian derived above. These two operators satisfy the
commutation relation $[\theta_{i}, L_{j}] = i\delta_{ij}$, and
thus the uncertainty relation $\Delta L_{i} \Delta \theta_{i}
\gtrsim 1$ works. Fluctuations of the $\theta_{i}$ field
correspond to bosonic collective excitations, here associated with
zero sound modes of a Fermi liquid when it becomes
condensed.\cite{Florens} This can be justified from the fact that
the dispersion of the $\theta_{i}$ field in its condensed phase is
given by that of sound waves.

The quantity $\varphi_{i}$ is the CDW order parameter in Eq. (2).
In the formulation presented in Eq. (5), however, it transforms as
the time component of a U(1) gauge field. Under the U(1) gauge
transformation for the matter fields, $f_{i\sigma} \rightarrow
e^{i\phi_{i}}f_{i\sigma}$ and $\theta_{i} \rightarrow \theta_{i} +
\phi_{i}$, the effective potential should be transformed into
$\varphi_{i} \rightarrow \varphi_{i} +
\partial_{\tau}\phi_{i}$. This gauge-field aspect of the order parameter is introduced
due to the mapping of Eq. (3), which involved the new phase degree
of freedom.

Integrating over the $L_{i}$ field, Eq. (5) reads \bqa && Z =
\int{D[f_{i\sigma},\theta_{i},\varphi_{i}]}e^{-\int{d\tau} L} ,
\nn&& L = \sum_{i\sigma}f_{i\sigma}^{*}(\partial_{\tau} - \mu -
i\varphi_{i})f_{i\sigma} -
t\sum_{ij\sigma}f_{i\sigma}^{*}e^{i(\theta_{i} -
\theta_{j})}f_{j\sigma} \nn && +
\frac{1}{4U}\sum_{i}(\partial_{\tau}\theta_{i} - \varphi_{i} )^{2}
. \eqa This expression is nothing but the slave-rotor
representation of the Hubbard model, obtained by FG in a different
fashion.\cite{Florens} It is clear that the CDW order parameter
appears to be the time component of a U(1) gauge field. This can
be understood by the fact that physics of the CDW order parameter
is an effective potential.

This effective Lagrangian should be considered to generalize the
LGW theoretical framework. If fluctuations of the $\theta_{i}$
fields are ignored, the resulting effective field theory belongs
to the LGW framework. However, as clearly demonstrated by FG,
$\theta_{i}$ fluctuations are mainly responsible for the
metal-insulator transition occurring in the Hubbard model at
half-filling. Keeping the $\theta_{i}$ fluctuations, the effective
field theory for the Mott transition is naturally given by a gauge
theory.\cite{LeeLee} In this respect the Mott transition should be
viewed beyond the LGW paradigm.

A standard treatment of the hopping term in Eq. (6) yields the
effective Lagrangian \bqa && L_{eff} =
t\sum_{<ij>}(\alpha_{ij}\beta_{ij}^{*} +
\beta_{ij}\alpha_{ij}^{*}) \nn && +
\sum_{i\sigma}f_{i\sigma}^{*}(\partial_{\tau} - \mu -
i\varphi_{i})f_{i\sigma} \nn && -
t\sum_{<ij>\sigma}(f_{i\sigma}^{*}\beta_{ij}^{*}f_{j\sigma} +
f_{j\sigma}^{*}\beta_{ij}f_{i\sigma}) \nn && +
\frac{1}{4U}\sum_{i}(\partial_{\tau}\theta_{i} - \varphi_{i} )^{2}
\nn && - t\sum_{<ij>}(e^{i\theta_{i}}\alpha_{ij}e^{-i\theta_{j}} +
e^{i\theta_{j}}\alpha_{ij}^{*}e^{-i\theta_{i}}) , \eqa where
$\alpha_{ij}$ and $\beta_{ij}$ are spinon and chargon hopping
order parameters, respectively.

A saddle point analysis results in the self-consistent equations
\bqa && -i\overline{\varphi}_{i} = -
i\langle\partial_{\tau}\theta_{i}\rangle +
2U\langle\sum_{\sigma}f_{i\sigma}^{*}f_{i\sigma}\rangle , \nn &&
\alpha_{ij} =
\langle\sum_{\sigma}{f}_{i\sigma}^{*}f_{j\sigma}\rangle , ~~~~~
\beta_{ij} = \langle{e}^{i\theta_{j}}e^{-i\theta_{i}}\rangle , \nn
&& \langle\sum_{\sigma}f_{i\sigma}^{*}f_{i\sigma}\rangle = 1 -
\delta , \eqa where $\delta$ is hole concentration.

Considering low energy fluctuations around this saddle point, one
can set $\alpha_{ij} = \alpha{e}^{ia_{ij}}$, $\beta_{ij} =
\beta{e}^{ia_{ij}}$ and $\varphi_{i} = \overline{\varphi}_{i} +
a_{i\tau}$, where $\alpha =
|\langle\sum_{\sigma}{f}_{i\sigma}^{*}f_{j\sigma}\rangle|$ and
$\beta = |\langle{e}^{i\theta_{j}}e^{-i\theta_{i}}\rangle|$ are
amplitudes of the hopping order parameters, and $a_{ij}$ and
$a_{i\tau}$ are spatial and time components of U(1) gauge fields.
Inserting these into Eq. (7), we find an effective U(1) gauge
theory for the Mott transition \bqa && L_{eff} = L_{0} + L_{f} +
L_{\theta} , \nn && L_{0} = 2tN\alpha\beta , \nn && L_{f} =
\sum_{i\sigma}f_{i\sigma}^{*}(\partial_{\tau} - \mu -
i\overline{\varphi}_{i} - ia_{i\tau})f_{i\sigma} \nn && -
t\beta\sum_{<ij>\sigma}(f_{i\sigma}^{*}e^{-ia_{ij}}f_{j\sigma} +
h.c.) , \nn && L_{\theta} =
\frac{1}{4U}\sum_{i}(\partial_{\tau}\theta_{i} -
\overline{\varphi}_{i} - a_{i\tau})^{2} \nn && -
2t\alpha\sum_{<ij>}\cos(\theta_{j} - \theta_{i} - a_{ij}) , \eqa
where $N$ is a total number of lattice sites. Eq. (9) is our
starting point for the metal-insulator
transition.\cite{Uncertainty}

In this effective gauge theory two important facts should be taken
into account since they discriminate the DCMT from the BCMT. One
is an effective chemical potential $\mu_{eff} = \mu +
i\overline{\varphi}_{i}$ in the spinon Lagrangian $L_{f}$.
Particle-hole symmetry at half filling causes the effective
chemical potential to vanish. On the other hand, away from half
filling the particle-hole symmetry is broken, resulting in a
nonzero chemical potential.

The other important feature is a Berry phase term arising from the
phase-fluctuation term in the chargon Lagrangian $L_{\theta}$ \bqa
&& S_{B} = - \sum_{i}
\int_{0}^{\beta}{d\tau}\frac{1}{2U}\overline{\varphi}_{i}\partial_{\tau}\theta_{i}
\nn && = - \sum_{i}
\int_{0}^{\beta}{d\tau}\Bigl(\frac{1}{2U}\langle\partial_{\tau}\theta_{i}\rangle\partial_{\tau}\theta_{i}
+i\langle\sum_{\sigma}f_{i\sigma}^{\dagger}f_{i\sigma}\rangle\partial_{\tau}\theta_{i}
\Bigr) . \eqa At half filling the Berry phase does not play any
roles because time reversal symmetry considered in this paper
leads to $\langle\partial_{\tau}\theta_{i}\rangle = 0$, and the
average occupation number of spinons is given by
$\langle\sum_{\sigma}f_{i\sigma}^{\dagger}f_{i\sigma}\rangle = 1$.
Inserting this into the expression of Berry phase, one obtains
$S_{B} = -
\sum_{i}\int_{0}^{\beta}{d\tau}i\partial_{\tau}\theta_{i} = -
2\pi{i}\sum_{i}q_{i}$, where $q_{i}$ is an integer representing an
instanton number, here a vortex charge. Thus, the contribution of
Berry phase to the partition function is nothing because of
$e^{-S_{B}} = 1$. Away from half filling the Berry phase action is
obtained to be $S_B = - 2\pi{i}\delta\sum_{i}{q}_{i}$ with modular
$2\pi$. This results in a complex phase factor to the partition
function, given by $Z = \sum_{Q}e^{2\pi{i}\delta{Q}}Z_{Q}$, where
$Q = \sum_{i}q_{i}$ is a total instanton number, and $Z_{Q}$, the
partition function for a fixed $Q$. The observation of Berry phase
gives the motivation for this paper. In this paper we investigate
how the effect of Berry phase makes the DCMT differ from the BCMT.

\section{Bandwidth-control Mott transition}

First, we discuss the BCMT. Zero effective chemical potential and
no Berry phase effect result in the following effective field
theory \bqa && L_{f} =
\sum_{i\sigma}f_{i\sigma}^{*}(\partial_{\tau} -
ia_{i\tau})f_{i\sigma} -
t\beta\sum_{<ij>\sigma}(f_{i\sigma}^{*}e^{-ia_{ij}}f_{j\sigma} +
c.c.) , \nn && L_{\theta} =
\frac{1}{4U}\sum_{i}(\partial_{\tau}\theta_{i} - a_{i\tau})^{2} -
2t\alpha\sum_{<ij>}\cos(\theta_{j} - \theta_{i} - a_{ij}) . \eqa
In the absence of U(1) gauge fluctuations this effective action
was intensively studied by FG.\cite{Florens} For the mean field
treatment FG utilized large $N$ generalization of the chargon
field, and derived the saddle point equations in Eq. (8) at half
filling. They found that there exists a critical $U/t$ for chargon
condensation.\cite{XY} In the case of $U/t > (U/t)_{c}$ chargons
are gapped, but spinons are massless. Existence of charge gap but
no spin gap corresponds to a spin liquid Mott insulator. In the
spin liquid there is no coherent quasiparticle peak at zero
energy, and only incoherent hump was found near the energy $\pm
U$. In the case of $U/t < (U/t)_{c}$ condensation of chargons
occurs, causing a coherent quasiparticle peak at zero energy in
the presence of incoherent hump near the energy $\pm U$. As a
result a correlated paramagnetic metal appears. Furthermore, FG
analyzed the saddle point equations near the Mott critical point
$(U/t)_{c}$, and obtained mean field critical exponents for the
charge gap and the quasiparticle weight. They also found that the
effective mass of quasiparticles does not diverge near the Mott
critical point owing to the spinon dispersion.

However, the mean field analysis of FG should be checked in the
presence of U(1) gauge fluctuations since instanton excitations of
U(1) gauge fields can cause confinement of spinons and chargons,
completely spoiling the mean field picture. In two space and one
time dimensions [$(2+1)D$] it is well known that static charged
matter fields are always confined owing to instanton
condensation.\cite{Polyakov} For the mean field picture of the
spin liquid and the Mott transition to be physically meaningful
beyond the mean field level, the stability of Eq. (11) should be
guaranteed against instanton excitations in the RG sense.

Recently, the present author examined deconfinement of fermions in
the presence of a Fermi surface.\cite{Kim_FS_DQCP} It has been
argued that the fermion Lagrangian $L_{f}$ in Eq. (11) has a
nontrivial charged fixed point\cite{FS_CFP,Altshuler} as the
quantum electrodynamics in $(2+1)D$ ($QED_3$) without a Fermi
surface.\cite{Hermele_QED3,Kim_QED3} The present author
investigated the stability of the charged critical point against
instanton excitations,\cite{Kim_FS_DQCP} following the strategy in
Ref. \cite{Hermele_QED3}. In the presence of a Fermi surface the
conductivity $\sigma_{f}$ of fermions is shown to play the similar
role as the flavor number $N$ of Dirac fermions in the
$QED_3$.\cite{Kim_FS_DQCP} Since the flavor number of Dirac
fermions is proportional to screening channels for the gauge
propagator, large flavors weaken gauge fluctuations in the
$QED_3$. In the same way the conductivity of fermions near the
Fermi surface determines strength of gauge fluctuations.
Remarkably, the charged fixed point is found to be stable against
instanton excitations when the fermion conductivity is
sufficiently large.\cite{Kim_FS_DQCP} This implies that the U(1)
gauge field can be considered to be noncompact. Eq. (11) can be a
stable theory against instanton excitations. In this respect the
spin liquid state can survive beyond the mean field level. But,
the spinons are not free particles any more owing to long range
gauge interactions, resulting in an algebraic behavior of the
spin-spin correlation function with an anomalous critical
exponent.\cite{Altshuler,YBKim}

The Mott transition beyond the mean field description is more
complex owing to the dissipative nature of gauge fluctuations.
Integrating over the spinons, one can obtain the effective action
for the chargon and gauge fields in the continuum limit \bqa &&
S_{eff} = \int{d\tau}{d^2r} \Bigl[
\frac{1}{4U}(\partial_{\tau}\theta - a_{\tau})^{2} -
2t\alpha\cos({\nabla}\theta - \mathbf{a}) \Bigr] \nn && +
\frac{1}{\beta}\sum_{\omega_{n}}\int{dq_{r}}\frac{1}{2}a_{\mu}(q_r,i\omega_{n})
D_{\mu\nu}^{-1}(q_r,i\omega_{n})a_{\nu}(-q_r,-i\omega_{n}) , \nn
\eqa where $D_{\mu\nu}(q_r,i\omega_{n})$ is the renormalized gauge
propagator, given by \bqa && D_{\mu\nu}(q_r,i\omega_{n}) =
\Bigl(\delta_{\mu\nu} -
\frac{q_{\mu}q_{\nu}}{q^2}\Bigr)D(q_r,i\omega_{n}) , \nn &&
D^{-1}(q_{r},i\omega_{n}) = D_{0}^{-1}(q_{r},i\omega_{n}) +
\Pi(q_r,i\omega_{n}) . \eqa Here $D_{0}^{-1}(q_{r},i\omega_{n}) =
(q_r^2 + \omega_{n}^2)/g^{2}$ is the bare propagator of the gauge
field given by the Maxwell gauge action, resulting from
integration of high energy fluctuations of spinons and chargons.
$g$ is an internal gauge charge of the spinon and chargon.
$\Pi(q_r,i\omega_{n})$ is the self-energy of the gauge field,
given by a correlation function of spinon charge (number)
currents. Since the current-current correlation function is
calculated in the noninteracting fermion ensemble, its structure
is well known\cite{Nagaosa,Tsvelik} \bqa && \Pi(q_r,i\omega_{n}) =
\sigma(q_{r})|\omega_{n}| + \chi{q}_{r}^{2} . \eqa Here the spinon
conductivity $\sigma(q_{r})$ is given by $\sigma(q_{r}) \approx
k_{0}/q_{r}$ in the clean limit while it is $\sigma(q_{r}) \approx
\sigma_{0} = k_{0}l$ in the dirty limit, where $k_{0}$ is of order
$k_{F}$ (Fermi momentum), and $l$ the spinon mean free path
determined by disorder scattering. The diamagnetic susceptibility
$\chi$ is given by $\chi \sim m_{f}^{-1}$, where $m_{f} \sim
(t\beta)^{-1}$ is the band mass of spinons. The frequency part of
the kernel $\Pi(q,i\omega_n)$ shows the dissipative propagation of
the gauge field owing to particle-hole excitations of spinons near
the Fermi surface.

Eq. (12) should be a starting point for the BCMT. In the study of
FG\cite{Florens} U(1) gauge fluctuations are ignored, and thus the
physical picture of the Mott transition should be modified. In the
absence of U(1) gauge fluctuations the transition falls into the
XY universality class. However, long range gauge interactions
alter the XY universality nature into the inverted-XY (IXY)
universality class if the Landau damping term in Eq. (14) is
ignored, and only the Maxwell kinetic energy of the gauge field is
taken into account.\cite{Kleinert} This means that if one
considers a critical exponent $\nu$ associated with the charge gap
$\Delta_{g} \sim |U - U_{c}|^{\nu}$ with the critical value
$U_{c}$, the critical exponent changes from $\nu_{XY}$ of the XY
transition to $\nu_{IXY}$ of the IXY transition. Damped gauge
interactions are expected to modify the IXY Mott
transition.\cite{Bose_metal}

Performing the duality transformation for the phase field in Eq.
(12), we obtain the dual vortex action \bqa && S_{v} =
\int{d\tau}{d^2r} \Bigl[|(\partial_{\mu} - ic_{\mu})\Phi|^{2} +
m_{v}^{2}|\Phi|^{2} + \frac{u_{v}}{2}|\Phi|^{4} \nn && +
U(\partial\times{c})_{\tau}^{2} +
\frac{1}{4t\alpha}(\partial\times{c})_{r}^{2} -
ia_{\mu}(\partial\times{c})_{\mu} \Bigr] \nn && +
\frac{1}{\beta}\sum_{\omega_{n}}\int{dq_{r}}\frac{1}{2}a_{\mu}(q_r,i\omega_{n})
D_{\mu\nu}^{-1}(q_r,i\omega_{n})a_{\nu}(-q_r,-i\omega_{n}) . \nn
\eqa Here $\Phi$ is a vortex field, and $c_{\mu}$ a vortex gauge
field. $m_{v}$ is a vortex mass, given by $m_{v}^{2} \sim
(U/t)_{c} - U/t$, and $u_{v}$ a phenomenologically introduced
parameter for local interactions between vortices. $(U/t)_{c}$ is
the critical strength of local interactions, associated with the
Mott transition in the mean field level.

In the dual vortex formulation the BCMT arises from controlling
the vortex mass as a function of the parameter $U/t$. In the case
of $m_{v}^{2} < 0$ ($U/t > (U/t)_{c}$) condensation of vortices
occurs, resulting in a Mott insulator of chargons. In the case of
$m_{v}^{2} > 0$ ($U/t < (U/t)_{c}$) vortices are gapped, implying
condensation of chargons, and a paramagnetic metal results.

Performing the Gaussian integration for the gauge field $a_{\mu}$,
we obtain the effective vortex action \bqa && Z_{v} =
\int{D[\Phi,c_{\mu}]} e^{- S_{v}} , \nn && S_{v} =
\int{d\tau}{d^2r} \Bigl[ |(\partial_{\mu} - ic_{\mu})\Phi|^{2} +
m_{v}^{2}|\Phi|^{2} + \frac{u_{v}}{2}|\Phi|^{4} \nn && +
U(\partial\times{c})_{\tau}^{2} +
\frac{1}{4t\alpha}(\partial\times{c})_{r}^{2} \Bigr] \nn && +
\int{d\tau}{d\tau_1}d^2rd^2r_1\frac{1}{2}c_{\mu}(r,\tau)K_{\mu\nu}(r-r_1,\tau-\tau_1)c_{\nu}(r_1,\tau_1)
, \nn \eqa where the renormalized gauge propagator
$K_{\mu\nu}(r-r_1,\tau-\tau_1)$ is given by in energy-momentum
space \bqa && K_{\mu\nu}(q_{r},i\omega_{n}) =
\Bigl(\delta_{\mu\nu} -
\frac{q_{\mu}q_{\nu}}{q^2}\Bigr)K(q_{r},i\omega_{n}) , \nn &&
K(q_r,i\omega_n) = \frac{q_{r}^2 + \omega_{n}^{2}}{(q_r^2 +
\omega_{n}^2)/g^{2} + \sigma(q_{r})|\omega_{n}| + \chi{q}_{r}^{2}}
\nn && \approx \frac{q_{r}^2 + \omega_{n}^{2}}{(q_r^2 +
\omega_{n}^2)/\overline{g}^{2} + \sigma(q_{r})|\omega_{n}|} , \eqa
where $\overline{g}$ is a redefined variable including the
susceptibility. In the following we consider dirty cases
characterized by $\sigma(q_{r}) = \sigma_{0}$.

Before we analyze Eq. (16) by using an RG method, we consider two
physical limits; one is $\sigma_{0} \rightarrow 0$ corresponding
to an insulator of spinons, and the other, $\sigma_{0} \rightarrow
\infty$ identified with a perfect metal of spinons. In the spinon
insulator the kernel $K(q_r,i\omega_n)$ becomes a constant value,
making vortex gauge fluctuations ($c_{\mu}$) gapped, thus ignored
in the low energy limit. This is because long range gauge
interactions ($a_{\mu}$) make it massive the low energy mode
(Goldstone mode) represented by the vortex gauge field, appearing
at high energies. The usual $\Phi^{4}$ action for the vortex field
is obtained. On the other hand, in the perfect spinon metal gauge
fluctuations $a_{\mu}$ are completely screened by spinon
excitations, causing the kernel to vanish, and the Maxwell gauge
action for the vortex gauge field results. The resulting vortex
action is reduced to the standard scalar QED$_{3}$. Varying the
spinon conductivity $\sigma_{0}$, these two limits would be
connected.

We perform an RG analysis for Eq. (16). Anisotropy in the Maxwell
gauge action for the vortex gauge field is assumed to be
irrelevant, and only the isotropic Maxwell gauge action is
considered by replacing $U, 1/4t\alpha$ with $1/(2e_{v}^{2})$.
Here $e_{v}$ is a vortex charge. In the limit of small anisotropy
the anisotropy was shown to be irrelevant at one loop
level.\cite{Herbut} To address the quantum critical behavior at
the Mott transition, we introduce the scaling $r = e^{l}r'$ and
$\tau = e^{l}\tau'$,\cite{Relativity} and consider the
renormalized theory at the transition point $m_{v}^{2} = 0$ \bqa
&& S_{v} = \int{d\tau'}{d^{D-1}r'} \Bigl[
Z_{\Phi}|(\partial'_{\mu} - ie_{v}c_{\mu})\Phi|^{2} +
Z_{u}\frac{u_{v}}{2}|\Phi|^{4} \nn && +
\frac{Z_{c}}{2}(\partial'\times{c})^{2} \Bigr] , \eqa where
$Z_{\Phi}$, $Z_{u}$ and $Z_{c}$ are the renormalization factors
defined by \bqa && \Phi =
e^{-\frac{D-2}{2}l}Z_{\Phi}^{\frac{1}{2}}\Phi_{r} \mbox{, } \mbox{
} c_{\mu} = e^{-\frac{D-2}{2}l}Z_{c}^{\frac{1}{2}}c_{\mu{r}} , \nn
&& e_{v}^2 = e^{-(4-D)l}Z_{c}^{-1}e_{vr}^{2} \mbox{,    } \mbox{ }
u_{v} = e^{-(4-D)l}Z_{u}Z^{-2}_{\Phi}u_{vr} . \eqa In the
renormalized action Eq. (18) the subscript $r$ implying
"renormalized" is omitted for simple notation.

Evaluating the renormalization factors at one loop level, the RG
equations are obtained to be \bqa && \frac{de_{v}^{2}}{dl} =
(4-D)e_{v}^{2} - \Bigl( \lambda{N}_{v} +
\frac{\zeta}{\sigma_{0}}\Bigr){e}_{v}^{4} , \nn &&
\frac{du_{v}}{dl} = (4-D)u_{v} +
h(\sigma_{0},e_{v}^{2})e_{v}^{2}u_{v} \nn && -
\rho(N_{v}+4){u}_{v}^{2} - g(\sigma_{0},e_{v}^{2})e_{v}^{4} . \eqa
Here $\lambda,\zeta,\rho$ are positive numerical constants, and
$h(\sigma_{0},e_{v}^{2}),g(\sigma_{0},e_{v}^{2})$ are analytic and
monotonically increasing functions of $\sigma_{0}$. $N_{v}$ is the
flavor number of the vortex field, here given by $N_{v} = 1$.

The first RG equation for the vortex charge can be understood in
the following way. Integrating out critical vortex fluctuations
near the critical point $m_{v}^{2} \sim 0$, we obtain the singular
contribution for the effective gauge action \bqa && S_{c} =
\frac{1}{\beta}\sum_{\omega_{n}}\int{d^{2}q_{r}} \frac{1}{2}
c_{\mu}(q_r,i\omega_n)\Xi_{\mu\nu}(q_r,i\omega_{n})c_{\nu}(-q,-i\omega_{n})
, \nn && \Xi_{\mu\nu}(q_r,i\omega_{n}) = \Bigl(\delta_{\mu\nu} -
\frac{q_{\mu}q_{\nu}}{q^2}\Bigr)\Xi(q_r,i\omega_{n}) , \nn &&
\Xi(q_r,i\omega_{n}) = \frac{N_{v}}{8}\sqrt{q_{r}^2 +
\omega_{n}^{2}} + K(q_{r},i\omega_{n}) \nn && \approx
\frac{N_{v}}{8}\sqrt{q_{r}^2 + \omega_{n}^{2}} + \frac{q_{r}^2 +
\omega_{n}^{2}}{\sigma_{0}|\omega_{n}|} . \nonumber \eqa The first
term in the kernel $\Xi(q_r,i\omega_{n})$ results from the
screening effect of the vortex charge via vortex polarization,
causing the $-\lambda{N}_{v}e_{v}^{4}$ term in the RG equation
while the second originates from that via spinon excitations,
yielding the $- ({\zeta}/{\sigma_{0}})e_{v}^{4}$ term. The first
$(4-D)e_{v}^{2}$ term denotes the bare scaling dimension of the
vortex charge.

For the second RG equation, unfortunately, we do not know the
exact functional forms of $h(\sigma_{0},e_{v}^{2})$ and
$g(\sigma_{0},e_{v}^{2})$ owing to the complexity of the gauge
kernel. Owing to the spinon contribution $K(q_{r},i\omega_{n})$
[Eq. (17)] the kernel of the gauge propagator ($c_{\mu}$) \bqa &&
D_{c}(q_{r},i\omega_{n}) = \frac{1}{q_{r}^{2} + \omega_{n}^{2} +
e_{v}^{2}K(q_{r},i\omega_{n})} \nn && \approx
\frac{\sigma_{0}|\omega_{n}|}{(q_{r}^{2} +
\omega_{n}^{2})(e_{v}^{2} + \sigma_{0}|\omega_{n}|)} \nonumber
\eqa should be utilized instead of the Maxwell propagator in
calculating one loop
diagrams.\cite{Kleinert,Herbut,Ye,Tesanovic_Herbut} Note the
dependence of the vortex charge $e_{v}^{2}$ in the effective gauge
propagator. This gives the dependence of the vortex charge to the
analytic functions $h(\sigma_{0},e_{v}^{2})$ and
$g(\sigma_{0},e_{v}^{2})$. Although the exact functional forms are
not known, the limiting values of these functions are clearly
revealed. In the limit of $\sigma_{0} \rightarrow 0$ the gauge
kernel vanishes, thus causing $h(\sigma_{0} \rightarrow
0,e_{v}^{2}) \rightarrow 0$ and $g(\sigma_{0} \rightarrow
0,e_{v}^{2}) \rightarrow 0$. In the small $\sigma_{0}$ limit the
gauge kernel is given by \bqa && D_{c}(q_{r},i\omega_{n}) \approx
\frac{\sigma_{0}}{e_{v}^{2}}\frac{|\omega_{n}|}{q_{r}^{2} +
\omega_{n}^{2}} , \nonumber \eqa thus resulting in
$h(\sigma_{0},e_{v}^{2}) = c_{h}\sigma_{0}/e_{v}^{2}$ and
$g(\sigma_{0},e_{v}^{2}) = c_{g}\sigma_{0}^{2}/e_{v}^{4}$, where
$c_{h}$ and $c_{g}$ are positive numerical constants. On the other
hand, in the limit of $\sigma_{0} \rightarrow \infty$ the gauge
kernel is reduced to the Maxwell one $D_{c}(q_{r},i\omega_{n}) =
1/(q_{r}^{2} + \omega_{n}^{2})$. Thus, $h(\sigma_{0} \rightarrow
\infty,e_{v}^{2}) \rightarrow c_{1}$ and $g(\sigma_{0} \rightarrow
\infty,e_{v}^{2}) \rightarrow c_{2}$ are obtained, where $c_{1}$
and $c_{2}$ are positive numerical constants.\cite{Kleinert,Ye} As
a result, Eq. (20) is reduced to the RG equation of the $\Phi^{4}$
theory\cite{Cubic} in the limit of $\sigma_{0} \rightarrow 0$ \bqa
&& \frac{du_{v}}{dl} = (4-D)u_{v} - \rho(N_{v}+4){u}_{v}^{2},
\nonumber \eqa and that of the scalar QED$_3$\cite{Kleinert,Ye} in
the limit of $\sigma_{0} \rightarrow \infty$, \bqa &&
\frac{de_{v}^{2}}{dl} = (4-D)e_{v}^{2} - \lambda{N}_{v}
{e}_{v}^{4} , \nn && \frac{du_{v}}{dl} = (4-D)u_{v} +
c_{1}e_{v}^{2}u_{v} - \rho(N_{v}+4){u}_{v}^{2} - c_{2}e_{v}^{4} .
\nonumber \eqa In the small $\sigma_{0}$ limit the RG equations
(20) result in \bqa && \frac{de_{v}^{2}}{dl} = (4-D)e_{v}^{2} -
\Bigl( \lambda{N}_{v} + \frac{\zeta}{\sigma_{0}}\Bigr){e}_{v}^{4}
, \nn && \frac{du_{v}}{dl} = (4-D)u_{v} + c_{h}\sigma_{0}u_{v} -
\rho(N_{v}+4){u}_{v}^{2} - c_{g}\sigma_{0}^{2} . \nonumber \eqa

In the scalar QED$_3$ there is a delicate issue about the
existence of the charged fixed point ($e_{v}^{*2} \not=
0$).\cite{Kleinert,Herbut,Tesanovic_Herbut} In this paper we do
not touch this issue. Instead we assume the existence of the
charged critical point in the scalar QED$_3$ by controlling the
$\lambda$ value. Then, the charged critical point
$(e_{v}^{*2}(\sigma_{0}),u_{v}^{*}[e_{v}^{*2}(\sigma_{0})])$ in
Eq. (20) is expected to vary as a function of the spinon
conductivity in the range of \bqa && e_{v}^{*2}(\sigma_{0}
\rightarrow 0) = 0 < e_{v}^{*2}(\sigma_{0}) <
e_{v}^{*2}(\sigma_{0} \rightarrow \infty) = 1/(\lambda{N}_{v}) ,
\nn && u_{v}^{*}[e_{v}^{*2}(\sigma_{0} \rightarrow \infty)] <
u_{v}^{*}[e_{v}^{*2}(\sigma_{0})] <
u_{v}^{*}[e_{v}^{*2}(\sigma_{0} \rightarrow 0)] , \nonumber \eqa
where the fixed point $(e_{v}^{*2}(\sigma_{0} \rightarrow
0),u_{v}^{*}[e_{v}^{*2}(\sigma_{0} \rightarrow 0)])$ corresponds
to the IXY one in the original boson model [Eq. (12)], and the
fixed point $(e_{v}^{*2}(\sigma_{0} \rightarrow
\infty),u_{v}^{*}[e_{v}^{*2}(\sigma_{0} \rightarrow \infty)])$
coincides with the XY one in Eq. (12). The spinon contribution
($\sigma_{0}$) connects the XY fixed point to the IXY one smoothly
in the chargon action Eq. (12).\cite{Critical_Theory} This implies
that the critical exponents near the Mott transition change
continuously, depending on the spinon conductivity. This would be
measured in some experiments. Because the spinon conductivity
depends on disorder, we would have some different critical points
by controlling density of disorder, resulting in various critical
exponents between the exponents of the XY and IXY transitions.
However, one interesting possibility should be taken into account
that the glassy behavior of the chargon field can originate from
random potentials. This important subject is under current
investigation.

\section{Doping-control Mott transition}

Next, we investigate the DCMT, described by the effective field
theory \bqa && L_{f} =
\sum_{i\sigma}f_{i\sigma}^{*}(\partial_{\tau} - \mu_{eff} -
ia_{i\tau})f_{i\sigma} \nn && -
t\beta\sum_{<ij>\sigma}(f_{i\sigma}^{*}e^{-ia_{ij}}f_{j\sigma} +
c.c.) , \nn && L_{\theta} =
\frac{1}{4U}\sum_{i}(\partial_{\tau}\theta_{i} - a_{i\tau})^{2} -
2t\alpha\sum_{<ij>}\cos(\theta_{j} - \theta_{i} - a_{ij}) \nn && +
i\delta(\partial_{\tau}\theta_{i} - a_{i\tau}) . \eqa Note the
presence of the effective chemical potential and Berry phase. This
Lagrangian is analyzed by employing a duality
transformation.\cite{Balents1,Tesanovic} In the dual formulation
the effect of Berry phase is represented as effective magnetic
fields for dual vortex variables.

Following the previous section, the duality transformation of the
chargon Lagrangian results in \bqa && {\cal L}_{v} =
|(\partial_{\mu} - ic_{\mu})\Phi|^{2} + m_{v}^{2}|\Phi|^{2} +
\frac{u_v}{2}|\Phi|^{4} - \overline{h}(\partial\times{c})_{\tau}
\nn && + U(\partial\times{c})_{\tau}^{2} +
\frac{1}{4t\alpha}(\partial\times{c})_{x}^{2} , \eqa where the
U(1) gauge field $a_{\mu}$ was ignored in the mean field level.
The Berry phase effect is reflected as an effective magnetic field
$\overline{h} = -2U\delta$ for the vortex field in the term
$-\overline{h}(\partial\times{c})_{\tau}$. Remember the expression
of the vortex mass $m_{v}^{2} \sim (U/t)_{c} - U/t$. A cautious
reader may suspect that the vortex mass should depend on hole
concentration. From the discussion below Eq. (9) it is important
to note that the effect of hole doping appears only in the
chemical potential and Berry phase terms. Furthermore, at half
filling Eq. (22) should be dual to the chargon Lagrangian
$L_{\theta}$ in Eq. (11). Thus, the vortex mass should depend on
only the parameter $U/t$.

In the dual vortex formulation the BCMT is driven by controlling
the vortex mass, as shown in the previous section. On the other
hand, the DCMT is nothing to do with the vortex mass. Instead,
controlling the effective magnetic field causes the Mott
transition. This leads us to consider that the nature of the DCMT
differs from the BCMT.

The presence of the effective magnetic field reminds us of a well
known Hofstadter problem for vortex fields. In the context of a
superfluid-insulator transition this was extensively studied in
Refs. \cite{Balents1,Tesanovic}. Here we briefly sketch the
procedure and key results. We first investigate the nature of a
doped Mott insulating state in a mean field fashion, i.e., the
absence of U(1) gauge fluctuations $a_{\mu}$, and discuss the DCMT
beyond the mean field level.

Following Ref. \cite{Balents1}, we consider commensurate hole
concentration $\delta = p/q$, where $p$ and $q$ are relatively
prime integers. Under this effective magnetic field $\delta$, the
vortex Lagrangian Eq. (22) has $q$-fold degenerate minima in the
magnetic Brillouin zone. Low energy fluctuations near the $q$-fold
degenerate vacua are assigned to be $\Psi_{l}$ with $l = 0, ...,
q-1$. A key question is how to construct a LGW free energy
functional in terms of the $\Psi_{l}$ fields. Constraints for an
effective potential of $\Psi_{l}$ are symmetry properties
associated with lattice translations and rotations in the presence
of the effective magnetic fields.\cite{Balents1} Based on the
symmetry properties one can construct a LGW free energy functional
of $\Psi_{l}$, and perform a standard mean field analysis. In this
free energy a superfluid of original bosons (chargons) is given by
$\langle\Psi_{l}\rangle = 0$ for all $l = 0, ..., q-1$ while a
Mott insulator is characterized by $\langle\Psi_{l}\rangle \not=
0$ for at least one $l$. Although the free energy functional has
all symmetries, the ground state can be symmetry-broken. In other
words, the Mott insulator can have broken translational
symmetries.

To see this, one can construct a density wave order parameter by
considering bilinear and gauge-invariant combinations of the low
energy vortices $\Psi_{l}$. Condensation of $\Psi_{l}$ leads to a
nonzero value of the density wave order parameter, causing a
vortex density wave. A density wave of vortices can be interpreted
as a crystalline phase of doped holes in the original
language.\cite{DHLee,Balents1,Tesanovic} In appendix we review the
simple $q = 2$ case. Combining this chargon physics with the
spinon physics, we can conclude that a doped Mott insulator
consists of a density wave of chargons and a spin liquid of
spinons with a Fermi surface. It should be noted that this doped
Mott insulator is different from the Mott insulator at half
filling because there is no charge order in the undoped Mott
insulator.

Remember that the crystalline phase of doped holes is nothing to
do with the spin liquid in the mean field level. Integrating out
the gapped chargon degrees of freedom, we obtain the same
spinon-gauge action for the doped spin liquid with that for the
undoped one beyond the mean field level. It has been argued that
the spinon-gauge action is a critical field theory at the
nontrivial charged fixed point,\cite{Kim_FS_DQCP} as discussed in
the previous section. Thus, the spin-spin correlation function
shows a power law behavior with respect to frequency and
temperature.\cite{Altshuler,YBKim} On the other hand, no infrared
response for charge fluctuations is expected owing to the Mott
gap. Instead, the charge order would be reflected in the electron
density of states as a spatially modulated pattern because of the
translational symmetry breaking.\cite{STM} The electron density of
states is proportional to the imaginary part of the electron green
function, given by convolution of the spinon and chargon
propagators in the slave-rotor formulation. Thus, the spatial
inhomogeneity of the chargon distribution results in the spatially
modulated pattern in the density of states. Because there is
excitation gap in the chargon spectrum, only incoherent hump would
be shown in the electron spectral function. This is another
different point from the usual density wave.

One cautious person may suspect the coexistence of the spin liquid
and charge density wave (CDW) because such a commensurate CDW can
destroy the spinon Fermi surface through a space-dependent
effective chemical potential, causing the spin liquid to be
unstable. However, we argue that the spinon Fermi surface can be
preserved even in the commensurate CDW when the Fermi surface
nesting is not perfect due to interaction or frustration effects.
Considering the low energy vortex excitations $\Psi_{l}$ near the
$q$-fold degenerate vacua, one can find the effective dual action
\bqa && S_{f} = \int {d\tau}\Bigl[
\sum_{i\sigma}f_{i\sigma}^{*}(\partial_{\tau} - \mu -
i\overline{\varphi}_{i} - ia_{i\tau})f_{i\sigma} \nn && -
t\beta\sum_{<ij>\sigma}(f_{i\sigma}^{*}e^{-ia_{ij}}f_{j\sigma} +
h.c.) \Bigr] -
\frac{1}{g^{2}}\sum_{\mu}\cos(\partial\times{a})_{\mu} , \nn &&
S_{v} = - t_{v}
\sum_{<nm>l}\Psi_{n}^{(l)*}e^{ic_{nm}}\Psi_{m}^{(l)} +
V(|\Psi_{n}^{(l)}|) \nn && -
\frac{1}{e_{v}^{2}}\sum_{\mu}\cos(\partial\times{c})_{\mu} +
i\sum_{<\mu\nu>}a_{\mu\nu}(\partial\times{c})_{\mu} . \eqa Here $l
= 1, ..., q$ corresponds to a color index of low energy vortex
fields, and $V(|\Psi_{n}^{(l)}|)$ is an effective vortex potential
determined by symmetry properties, where the coefficients are
effectively doping dependent (see appendix). The last gauge action
in the spinon sector originates from high energy contributions of
matter fields, where $g$ is an internal gauge charge of spinons.

The question is what happens in the Fermi surface when vortex
condensation occurs, resulting in translational symmetry breaking.
Ignoring spinon-gauge fluctuations $a_{\mu\nu}$ as the mean field
approximation, the spinon-gauge action is completely decoupled
from the vortex-gauge action, as discussed before. This indicates
that the Fermi surface is not affected by the CDW formation in the
vortex sector. Now, we allow spinon-gauge excitations. Integrating
out $a_{i\tau}$ in the limit of $g \rightarrow \infty$, one
obtains the constraint \bqa && (\nabla\times{c})_{i} =
\sum_{\sigma}{f_{i\sigma}^{\dagger}f_{i\sigma}} . \nonumber \eqa
When the vortices are condensed to cause translational symmetry
breaking, the above quantity should depend on positions. This
effect can be introduced in the spinon action by allowing a
position-dependent effective chemical potential, interpreted as a
higher order effect due to gauge fluctuations. In this case the
commensurate CDW can destroy the Fermi surface. However, it should
be noted that this depends on the shape of the Fermi surface. When
the perfect nesting of the Fermi surface does not appear due to
interaction or frustration effects, only partial parts of the
Fermi surface would open the CDW gap, and other parts of the Fermi
surface, not connected by the CDW wave vector, are expected to
remain gapless. This would indeed happen when there is
frustration, destroying the Fermi surface nesting. This
expectation coincides with our ignorance of spin ordering because
strong frustration kills magnetic ordering.

We should emphasize that the above discussion is applied to the $g
\rightarrow \infty$ limit. Since we are considering low energy
fluctuations, high energy matter fields should be integrated out,
resulting in the Maxwell gauge action, the last term in $S_{f}$ of
Eq. (23). Then, the above constraint cannot be used because one
cannot integrate out $a_{i\tau}$ directly. In this case of
spin-charge separation due to a finite value of $g$, the
position-dependent dual magnetic flux is not directly related with
the spinon density owing to the spinon-gauge flux. If one utilizes
average values in the constraint equation instead of the operator
identity, the same argument above can be applied. Generically,
there should be a gapless Fermi surface, not connected by the CDW
wave vector, at least in the frustrated lattice.

The present doped spin liquid is an interesting new phase in the
respect that a conventional order described by the CDW order
parameter (in the LGW paradigm) and an exotic order associated
with a conserved internal gauge
flux\cite{Senthil_deconfinement,Kim_deconfinement} coexist. This
phase is expected to be stable beyond the mean field level because
instanton excitations can be suppressed via spinon excitations
near the Fermi surface, as discussed before. A next important
question is which phase this doped spin liquid evolves into. More
concretely, when chargon condensation occurs, does the CDW order
survive? Remember that in the BCMT chargon condensation results in
a Fermi liquid, where condensed chargons are confined with spinons
to form electrons (quasiparticles). This corresponds to a
Higgs-confinement phase in the context of gauge theory, where the
internal gauge flux is not
conserved.\cite{Senthil_deconfinement,Kim_deconfinement} In the
DCMT the chargon condensation also causes electronic
quasiparticles. In this case the CDW order is expected to
disappear, thus resulting in the same Fermi liquid as that in the
BCMT. See Eq. (22). Because we are considering gapped vortex
excitations, they can be ignored in the low energy limit. Thus,
there remain uniform effective magnetic fields. This implies that
condensed chargons are homogenously distributed. As a result the
CDW order of chargons disappears. Another way to say this is that
since there are no vortex charges (owing to the gap in the vortex
excitations) in the Berry phase term Eq. (10), the effect of Berry
phase disappears. This is analogous to the case in the nonlinear
$\sigma$ model for the quantum antiferromagnet, where the Berry
phase effect can be ignored in the antiferromagnetic
phase.\cite{Senthil_deconfinement}

Now we discuss a critical field theory for the DCMT. Integrating
over the spinon and gauge ($a_{\mu}$) excitations in Eq. (23) as
performed at half filling, one can construct the following
effective field theory \bqa && S_{eff} = \int{d\tau}d^2r\Bigl[
\sum_{l = 0}^{q-1}|(\partial_{\mu} - ic_{\mu})\Psi_{l}|^{2} +
V(\Psi_{l}) \nn && - (\overline{h} -
\overline{h}_{q})(\partial\times{c})_{\tau} +
U(\partial\times{c})^{2}_{\tau} +
\frac{1}{4t\alpha}(\partial\times{c})_{x}^{2} \Bigr] \nn && +
\int{d\tau}{d\tau_1}d^2rd^2r_1\frac{1}{2}c_{\mu}(r,\tau)K_{\mu\nu}(r-r_1,\tau-\tau_1)c_{\nu}(r_1,\tau_1)
. \nn \eqa $K_{\mu\nu}$ results from the anomalous contribution of
spinon-gauge fluctuations to vortex-gauge excitations, given by
Eq. (17) with a different $\sigma_{0}$ owing to the chemical
potential $\mu_{eff}$. $\overline{h} = -2U\delta$ is an applied
effective magnetic field, and $\overline{h}_{q} = -2U\delta_{q}$ a
nearby one with commensurate hole concentration $\delta_{q} =
p/q$. One can estimate a critical effective magnetic field
$\overline{h}_{c}$ with a given $U/t > (U/t)_{c}$ by calculating
the condensation energy. The critical hole concentration
$\delta_{c}$ corresponding to the critical magnetic field
$\overline{h}_{c}$ would be different from $\delta_{q}$ generally.
In this case one may determine a moderate value of $q$ near the
critical doping $\delta_{c}$. Then, there remain residual
effective magnetic fields $\overline{h} - \overline{h}_{q}$,
corresponding to the incommensurability $\delta - \delta_{q}$.

We propose that Eq. (24) is a starting point for the DCMT. If the
vortex "superconductor" falls into the type-I class, the residual
magnetic field would be expelled owing to the dual "Meissner"
effect. A critical field theory for this Mott transition is
expected to be without the residual magnetic field \bqa && S_{eff}
= \int{d\tau}d^2r\Bigl[ \sum_{l = 0}^{q-1}|(\partial_{\mu} -
ic_{\mu})\Psi_{l}|^{2} + V(\Psi_{l}) \nn && +
U(\partial\times{c})^{2}_{\tau} +
\frac{1}{4t\alpha}(\partial\times{c})_{x}^{2} \Bigr] \nn && +
\int{d\tau}{d\tau_1}d^2rd^2r_1\frac{1}{2}c_{\mu}(r,\tau)K_{\mu\nu}(r-r_1,\tau-\tau_1)c_{\nu}(r_1,\tau_1)
. \nn \eqa Because the effective vortex action depends on $q$ and
$V(\Psi_{l})$, it is difficult to predict critical vortex dynamics
for general $q$ values. The $q = 1$ case corresponds to the
undoped spin liquid, already discussed in the previous section. In
the $q = 2$ case the effective vortex potential is obtained to be
\bqa && V(\psi_{l}) = m^{2}(|\psi_{0}|^{2} + |\psi_{1}|^{2}) +
u_{4}(|\psi_{0}|^{2} + |\psi_{1}|^{2})^{2} \nn && +
v_{4}|\psi_{1}|^{2}|\psi_{2}|^{2} -
v_{8}[(\psi_{1}^{*}\psi_{2})^{4} + H.c.] , \nonumber \eqa well
discussed in appendix. At the critical point $m^{2} = 0$ the last
eighth-order term is certainly irrelevant owing to its high order.
Furthermore, the cubic anisotropy term ($v_{4}$) is well known to
be irrelevant in the case of $q < q_{c} = 4$.\cite{Cubic} As a
result, the Heisenberg fixed point ($v_{4}^{*} = 0$ and $u_{4}^{*}
\not= 0$) appears in the absence of vortex gauge
fluctuations.\cite{Cubic} Introducing the vortex gauge fields at
the Heisenberg fixed point, we have qualitatively the same fixed
point with Eq. (16) except the $q = 2$ vortices. Since the charged
critical point depends on the spinon conductivity, the critical
exponents vary as a function of the spinon conductivity. At higher
$q$ values we do not understand the nature of the Mott transition
owing to the complexity of the vortex potential. Generally
speaking, a continuous Mott transition from the U(1) spin liquid
with a commensurate density wave order to the Fermi liquid is
possible.

On the other hand, if the vortex superconductor belongs to the
type-II class, the residual magnetic field can penetrate the
superconductor, forming a dual Abrikosov "vortex" lattice. This
corresponds to an incommensurate Mott insulator, where hole
density is $\delta \not= \delta_{q}$.\cite{Balents1} In this case
the nature of the Mott transition from the U(1) spin liquid with
an incommensurate density wave order to the Fermi liquid is not
clear owing to the Berry phase effect. Furthermore, the Landau
damping term should be taken into account in the critical field
theory as the case of the BCMT because it changes the nature of
the Mott transition. A continuous transition to the Fermi liquid
may be possible in this case. A detailed analysis of this DCMT is
beyond the scope of this paper.

\begin{figure}
\includegraphics[width=8cm]{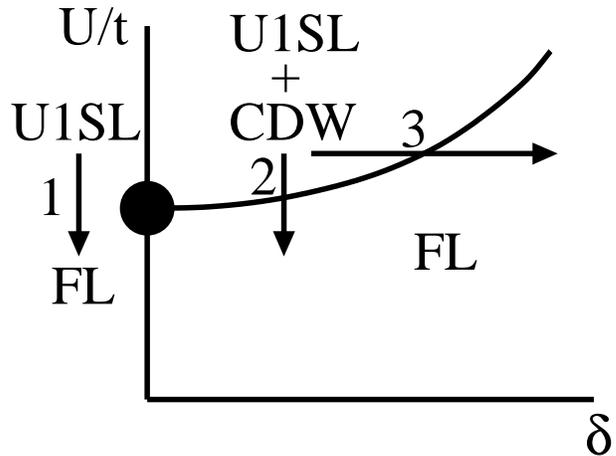}
\caption{\label{Fig. 1} A schematic phase diagram in the
slave-rotor representation of the Hubbard model}
\end{figure}

We propose a phase diagram Fig. 1 in the slave-rotor
representation of the Hubbard model on two dimensional square
lattice. Here U1SL, FL, and CDW represent U(1) spin liquid, Fermi
liquid, and charge density wave, respectively.

The route 1 is the BCMT from U1SL to FL at half filling while the
route 2, that from U1SL + CDW to FL at commensurate hole
concentration. In these cases we showed critical field theories,
and discussed nature of the continuous phase transitions.

The route 3 is the DCMT at a finite $U/t$. In this case the
critical field theory depends on the nature of the vortex
superconductor. Nature of the quantum phase transition from the
U1SL with an incommensurate CDW to the FL is not clear owing to
incommensurability.

\section{Discussion and summary}

So far, we considered a zero flux state, and thus obtained the
U(1) spin liquid with a Fermi surface. The uniform spin liquid
phase turns out to be stable (with respect to the flux phase) in
the triangular lattice at the mean-field level near the undoped
Mott transition point\cite{LeeLee}. However, it is important to
consider a $\pi$ flux phase since this phase is usually obtained
as a stable mean field state in the square lattice without
frustration\cite{Lee_review}. In the $\pi$ flux phase low energy
spinon excitations are given by Dirac fermions near four nodal
points. As a result the effective spinon Lagrangian is obtained to
be $QED_3$. For the BCMT at half filling there is no dissipation
in gauge fluctuations. The role of massless Dirac fermions is to
weaken gauge interactions, resulting from screening of gauge
charges owing to particle-hole polarization. Thus, increasing the
flavor number $N$ of Dirac fermions in the $1/N$ approximation,
the IXY transition is expected to turn into the XY
one\cite{QED_phi4}. On the other hand, for the DCMT there remains
dissipation in gauge fluctuations owing to a nonzero effective
chemical potential. Thus, damped gauge fluctuations would still
play some special roles in the DCMT.

In this paper we discussed how the doping-control Mott transition
differs from the bandwidth-control one based on the slave-rotor
representation of the Hubbard model. We found that the doped Mott
insulator consists of a crystalline phase of doped holes and a
U(1) spin liquid with a Fermi surface while the Mott insulator at
half filling is the U(1) spin liquid without any charge orders.
This originates from the fact that hole doping causes a Berry
phase term to the chargon field. This Berry phase effect results
in an effective magnetic field to the vortex field of the chargon
field. In the dual vortex formulation we showed that the
bandwidth-control Mott transition is driven by the sign change of
the vortex mass while the doping-control one is achieved by the
control of the effective magnetic field. The presence of dual
effective magnetic fields leads to translational symmetry breaking
when vortices are condensed. We argued that this charge order does
not destroy the spinon Fermi surface when there is strong
frustration, causing the Fermi nesting to disappear. As a result,
the spin liquid phase can remain stable to coexist with the
density wave. Furthermore, we pointed out that damped U(1) gauge
fluctuations resulting from spinon excitations should be taken
into account for both Mott transitions because the nature of the
Mott transitions is modified by the dissipative gauge excitations.
Performing a renormalization group analysis, we showed that the
Mott critical point depends on the spinon conductivity
characterizing the strength of dissipation. This interesting
result leads us to predict that varying the density of disorder
would cause different critical exponents because disorder
determines the conductivity of spinons.

\section{Acknowledgement}

K.-S. Kim especially thanks Prof. J.-H. Han for helpful
discussions.

\appendix*

\section{Mean field analysis}

In this appendix we review a mean field analysis for the $q = 2$
case following Refs. \cite{Fisher_Senthil,Kim_Berry}. Ignoring
vortex gauge fluctuations in Eq. (22), we can write down the
vortex action with the effective magnetic field $\delta = 1/2$ in
a lattice version \bqa && S_{M} = \int{d\tau} \Bigl[
\sum_{n}|\partial_{\tau}\Phi_{n}|^{2} -
\sum_{nm}\Phi_{n}^{\dagger}t^{v}_{nm}\Phi_{m} + V(|\Phi|) \Bigr] ,
\nn \eqa where $n, m$ label sites in the dual lattice, and the
sign of the hopping integral $t^{v}_{nm}$ around a plaquette is
$-1$ owing to the $\pi$ flux background.

The vortex hopping term can be easily diagonalized in the
eigenvectors $\chi_{n}^{0} = (1+\sqrt{2}) - e^{i\pi{n_{y}}}$ and
$\chi_{n}^{0} = e^{i\pi{n_{x}}}[(1+\sqrt{2}) + e^{i\pi{n_{y}}}]$,
resulting in two low energy modes near the momentum $k = (0, 0)$
and $k = (\pi, 0)$. Then, the low energy dynamics of this system
can be described by the low energy vortex fields $\Psi_{0n}$ and
$\Psi_{1n}$ in $\Phi_{n} = \Psi_{0n}\chi_{n}^{0} +
\Psi_{1n}\chi_{n}^{\pi}$.

In order to construct the effective vortex potential one can
introduce the following two complex fields \bqa && \psi_{0} =
\Psi_{0} + i\Psi_{1} , ~~~~~ \psi_{1} = \Psi_{0} - i\Psi_{1} .
\eqa Then, the symmetry transformations are given by \bqa &&
T_{x}: \psi_{0} \longrightarrow \psi_{1} , ~~~~~ \psi_{1}
\longrightarrow \psi_{0} , \nn && T_{y}: \psi_{0} \longrightarrow
i\psi_{1} , ~~~~~ \psi_{1} \longrightarrow -i\psi_{0} , \nn &&
R_{\pi/2}: \psi_{0} \longrightarrow e^{i\pi/4}\psi_{0} , ~~~~~
\psi_{1} \longrightarrow e^{-i\pi/4}\psi_{1} , \eqa where
$T_{x(y)}$ and $R_{\pi/2}$ are associated with lattice
translations and rotations. The LGW effective potential allowed by
these symmetry operations is obtained to be\cite{Fisher_Senthil}
\bqa && V(\psi_{l}) = m^{2}(|\psi_{0}|^{2} + |\psi_{1}|^{2}) +
u_{4}(|\psi_{0}|^{2} + |\psi_{1}|^{2})^{2} \nn && +
v_{4}|\psi_{0}|^{2}|\psi_{1}|^{2} -
v_{8}[(\psi_{0}^{*}\psi_{1})^{4} + H.c.] , \eqa where $m^{2}$ is
an effective vortex mass, $u_{4}$ a local interaction, $v_{4}$ the
cubic anisotropy, and $v_{8}$ breaking the U(1) phase
transformation $\psi_{0(1)} \rightarrow
e^{i\varphi_{0(1)}}\psi_{0(1)}$.

One cautious reader may ask how the coefficients in the LGW free
energy functional can be determined. Although the symmetry
constraints restrict the functional form of the effective
potential, they cannot determine the remaining parameters in the
free energy. Our question is whether these parameters are doping
dependent or not. Remember that there is no doping dependence in
the original vortex mass $m_{v}^{2}$ in Eq. (15). It depends on
only the parameter $U/t$. However, the effective parameters for
the low energy vortex fields should be considered to be doping
dependent. Consider a vortex vacuum resulting from large effective
magnetic fields $\delta$ in spite of $m_{v}^{2} < 0$ ($U/t >
(U/t)_{c}$), corresponding to chargon condensation. Decreasing
hole concentration, the flavor number $q$ of low energy vortices
would increase. Decreasing hole concentration further, some
components of the low energy vortices are expected to be
condensed. In this respect the coefficients in the LGW free energy
of $\Psi_{l}$ (or $\psi_{l}$) can be considered to depend on hole
concentration effectively.

Based on the effective vortex potential Eq. (A4), one can perform
a mean field analysis. Condensation of vortices occurs in the case
of $m^{2} < 0$ and $u_{4} > 0$. The signs of $v_{4}$ and $v_{8}$
then determine the ground state. For $v_{4} < 0$, both vortices
have a nonzero vacuum expectation value $|\langle\psi_{0}\rangle|
= |\langle\psi_{1}\rangle| \not= 0$, and their relative phase is
determined by the sign of $v_{8}$. In the case of $v_{8} > 0$ the
resulting vortex state corresponds to the columnar dimer order,
breaking the rotational and translational symmetries. In the case
of $v_{8} < 0$ the resulting phase exhibits the plaquette pattern,
braking the rotational symmetries. On the other hand, if $v_{4}
> 0$, the ground states are given by either
$|\langle\psi_{0}\rangle| \not= 0, |\langle\psi_{1}\rangle| = 0$
or $|\langle\psi_{0}\rangle| = 0, |\langle\psi_{1}\rangle| \not=
0$, and the sign of $v_{8}$ is irrelevant. In this case an
ordinary charge density wave order at wave vector $(\pi, \pi)$ is
obtained, breaking the translational symmetries.


\begin{thebibliography}{9}
\bibitem{Uncertainty} One cautious person may ask where the
conjugate partner of the CDW order parameter exists. Remember that
the order parameter plays the same role as the time component of a
gauge field. In this respect the conjugate partner of the order
parameter would be associated with the spatial components of the
gauge field. There is an uncertainty relation between electric
fields $e_{ij}$ and electromagnetic vector potentials $a_{ij}$,
$\Delta{e}_{ij}\Delta{a}_{ij} \gtrsim 1$. Note that the spatial
components of the gauge field result from the HS fields after
performing the HS transformation of the hopping term. In Eq. (9)
all quantum variables have their conjugate partners.
\bibitem{XY} This critical point should exist because the
effective Lagrangian of the chargon field corresponds to the
quantum XY model in the mean field level (without gauge
fluctuations).
\bibitem{Relativity} We note that the relativistic invariance is
preserved at the critical point despite the spinon contribution,
indicated by the dispersion relation $\omega_{c} \sim q_{r}$ for
the vortex gauge field. The dispersion relation can be obtained
from the renormalized propagator $\Xi(q_{r},i\omega_{n})$ in the
text.
\bibitem{Critical_Theory} The existence of the nontrivial
charged fixed point can be easily understood from the gauge action
at the critical point in the chargon action Eq. (12). The gauge
action can be written in a highly schematic form from Eq. (12) at
the critical point \bqa && {S}_{g} =
\frac{1}{\beta}\sum_{\omega_{n}}\int{d^{2}q_{r}}
\frac{1}{2}\mathbf{a}^{T}(q_{r},i\omega_{n})\Pi(q_{r},i\omega_{n})\mathbf{a}^{T}(-q_{r},-i\omega_{n})
, \nonumber \eqa where $\mathbf{a}^{T}(q_{r},i\omega_{n})$ is the
transverse vector of the gauge field. The gauge kernel is given by
\bqa && \Pi(q_{r},i\omega_{n}) =
\frac{N_{\theta}}{8}\sqrt{q_{r}^{2}+\omega_{n}^{2}} +
\sigma_{0}|\omega_{n}| , \nonumber \eqa where $N_{\theta}$ is the
flavor number of the chargon field. The first term results from
critical chargon fluctuations while the second one originates from
gapless spinon excitations near the Fermi surface. This gauge
action can be easily checked to be scale-invariant at the tree
level, giving the IXY fixed point in the $\sigma_{0} \rightarrow
0$ limit and the XY one in the $\sigma_{0} \rightarrow \infty$
limit. Thus, the spinon contribution characterized by the spinon
conductivity $\sigma_{0}$ connects these two fixed points
smoothly.
\end{thebibliography}
\end{document}